\documentclass[sigconf]{acmart}

\renewcommand\footnotetextcopyrightpermission[1]{}
\usepackage{hyperref}
\usepackage{cleveref}
\usepackage{url}
\usepackage{algorithm}
\usepackage{algpseudocode}
\usepackage{algorithmicx}
\AtBeginDocument{%
  \providecommand\BibTeX{{%
    \normalfont B\kern-0.5em{\scshape i\kern-0.25em b}\kern-0.8em\TeX}}}

\acmConference[]{}{}{}
\acmPrice{15.00}
\acmISBN{978-1-4503-XXXX-X/18/06}

\begin{document}


\title[]{SNEAP: A Fast and Efficient Toolchain for Mapping Large-Scale Spiking Neural Network onto NoC-based Neuromorphic Platform}


\author{Shiming Li}
\affiliation{%
  \institution{National University of Defense Technology}
  \city{Changsha}
  \country{China}}
\email{lishiming15@nudt.edu.cn}

\author{Shasha Guo}
\affiliation{%
  \institution{National University of Defense Technology}
  \city{Changsha}
  \country{China}}
\email{guoshasha13@nudt.edu.cn}

\author{Limeng Zhang}
\affiliation{%
  \institution{National University of Defense Technology}
  \city{Changsha}
  \country{China}}
\email{zhanglimeng@nudt.edu.cn}

\author{Ziyang Kang}
\affiliation{%
  \institution{National University of Defense Technology}
  \city{Changsha}
  \country{China}}
\email{kangziyang14@nudt.edu.cn}

\author{Shiying Wang}
\affiliation{%
  \institution{National University of Defense Technology}
  \city{Changsha}
  \country{China}}
\email{wangshiying18@nudt.edu.cn}

\author{Wei Shi}
\affiliation{%
  \institution{National University of Defense Technology}
  \city{Changsha}
  \country{China}}
\email{shiwei@nudt.edu.cn}

\author{Lei Wang}
\affiliation{%
  \institution{National University of Defense Technology}
  \city{Changsha}
  \country{China}}
\email{leiwang@nudt.edu.cn}

\author{Weixia Xu}
\affiliation{%
  \institution{National University of Defense Technology}
  \city{Changsha}
  \country{China}}
\email{xuweixia@nudt.edu.cn}


\begin{abstract}
  Spiking neural network (SNN), as the third generation of artificial neural networks, has been widely adopted in vision and audio tasks. Nowadays, many neuromorphic platforms support SNN simulation and adopt Network-on-Chips (NoC) architecture for multi-cores interconnection. 
  However, interconnection brings huge area overhead to the platform. Moreover, run-time communication on the interconnection has a significant effect on the total power consumption and performance of the platform. In this paper, we propose a toolchain called SNEAP (\underline{S}piking \underline{NE}ural network m\underline{AP}ping toolchain) for mapping SNNs to neuromorphic platforms with multi-cores, which aims to reduce the energy and latency brought by spike communication on the interconnection. 
  
  SNEAP includes two key steps: partitioning the SNN to reduce the spikes communicated between partitions, and mapping the partitions of SNN to the NoC to reduce average hop of spikes under the constraint of hardware resources. SNEAP can reduce more spikes communicated on the interconnection of NoC and spend less time than other toolchains in the partitioning phase. Moreover, the average hop of spikes is reduced more by SNEAP within a time period, which effectively reduces the energy and latency on the NoC-based neuromorphic platform. 
  
  The experimental results show that SNEAP can achieve 418$\times$ reduction in end-to-end execution time, and reduce energy consumption and spike latency, on average, by 23\% and 51\% respectively, compared with SpiNeMap.
  
\end{abstract}





\maketitle

\section{Introduction}

Spiking neuron networks (SNN) \cite{maass1997networks} is the third generation of artificial neural network (ANN) inspired by brain science.
At present, SNNs are widely adopted in image classification, pattern recognition tasks and so on \cite{diehl2015unsupervised}. A neuron in SNN accepts stimulus and generates spikes if its membrane potential exceeds the firing threshold. Neurons communicate with each other by spikes. Compared with current popular ANNs, SNNs have more biological characteristics and require lower power consumption when simulated with neuromorphic platforms \cite{diehl2016conversion}.

Neuromorphic platforms are gaining more attention recently. The typical examples are IBM’s TrueNorth \cite{akopyan2015truenorth}, Intel’s Loihi \cite{davies2018loihi}, ETH’s Dynapse \cite{moradi2017scalable}, UM’s SpiNNaker \cite{furber2012overview} etc. All of these neuromorphic platforms are based on Network-on-Chips (NoC) to connect multiple neuromorphic cores. In each neuromorphic core, there are fixed amounts of neurons. 

Mapping SNNs to various neuromorphic platforms is a key step in the application of neuromorphic platform. The general solution is dividing a SNN into multiple partitions, and then mapping these partitions to the neuromorphic cores. The neurons of each partition should not exceed the capacity of a single neuromorphic core. If these partitions cannot be mapped at once when the partitions outnumber the cores, multiple rounds of mapping are required to ensure that all partitions have been mapped and executed.

There are some mapping methods for deploying SNN to these neuromorphic platforms, such as PACMAN \cite{galluppi2012hierachical}, NEUTRAMS \cite{ji2016neutrams}, SCO \cite{lee2019system}, SpiNeMap \cite{balaji2019mapping}, and etc. But these mapping methods have some problems. PACMAN only partitions the SNN model and then sequentially maps the result of partitioning to the ARM cores, which leads to spike congestion on the NoC. SCO adopts sequential mapping methods, which minimize the neuromorphic cores usage to reduce the overhead of hardware resources. But this method does not optimizes the spikes communication between cores, resulting in increased spike latency and power consumption. Although SpiNeMap uses a two-stage optimization method to reduce the power consumption and latency of the neuromorphic platform, the entire process will take a huge amount of time for large-scale SNNs. Meanwhile, limited by the algorithm, SpiNeMap does not search out the best mapping scheme.

There are two challenges for SNN mapping. 
The first comes from the partitioning process. It is slow to partition the SNNs and hard to find the best solution with the minimized spike communications for larger SNNs. The second comes from the mapping process. A fast and efficient search algorithm needed to be proposed to find out the best mapping scheme that minimizes the spike latency and energy of the NoC-based neuromorphic platform. During the mapping process, the search algorithm continuously evaluates the metrics, such as average hop, latency, and energy. However, the evaluation of these metrics often requires to use real hardware or hardware simulator, which leads to a lot of time consumption and makes the entire optimization process unacceptable. 

To confront these challenges, we propose a toolchain for mapping a large-scale SNN onto a NoC-based neuromorphic platform, called SNEAP (\underline{S}piking \underline{NE}ural network m\underline{AP}ping toolchain). The toolchain includes four parts: profiling, partitioning, mapping, and evaluation. We first profile the connection information and the spike traces of a SNN from the software simulator. Then we use a multi-level graph partitioning method to quickly reduce the number of inter spike communications under the constraints of hardware structure. Subsequently, a  heuristic algorithm that selected from three algorithm are used to map partitions to the NoC architecture to optimize latency and energy. Finally, the mapping scheme is evaluated by NoC-based hardware simulator, Noxim++ \cite{balaji2019mapping}, so as to get key performance statistics.

Our contributions of this paper as follows:
\begin{itemize}
  \item We propose a toolchain to map SNN to underlying NoC-based neuromorphic platform. During the mapping process, average neuron communication latency and power consumption is minimalized.
  \item For large-scale SNN, we use an effective graph partitioning method to improve the quality of partitioning while reducing the partitioning time dramatically.
  \item We use an optimization algorithm that selected from three algorithms to minimize average neuron communication latency and power consumption during the mapping phase of the toolchain. 
  \item Average hop is used to evaluate the average neuron communication latency and power consumption instead of using the simulator to improve the search speed.
\end{itemize}

We evaluate SNEAP using several SNNs. The experiment result shows that SNEAP can achieve 418$\times$ reduction in end-to-end execution time, and reduce average energy consumption by 23\% and average spike latency by 51\%, compared to SpiNeMap \cite{balaji2019mapping}.

\section{Background \& Related works}
\subsection{NoC of Neuromorphic Platforms}
The neuromorphic platform aims at developing VLSI systems to mimic the neuro-biological networks of the nervous system - SNN. It is a large-scale parallel system composed of a large number of computing units called neuromorphic cores interconnected by NoC. NoC is responsible for managing communication in the neuromorphic platform. NoC structure generally uses a dimensional-order routing strategy to avoid deadlocks. According to the topology of NoC, two types of NoC are commonly used: NoC-tree and NoC-mesh. Examples include the NoC-mesh for TrueNorth and Loihi, multi-stage NoC-mesh for Dynapse \cite{moradi2017scalable}, and NoC-tree for CxQuad. 

SpiNNaker \cite{furber2012overview} simulates the brain by connecting 1 million ARM processors together in real-time. Eighteen ARM processors are integrated into one chip multiprocessor (CMP), and 216 CMPs form a complete system with a 2D toroidal mesh structure. Dynapse \cite{moradi2017scalable} is an advanced mixed-signal multi-core neuromorphic processor. Dynapse hse 4 cores, each core has 256 analog circuit neurons. These 256 analog neurons are placed on a 16x16 2D-Mesh. The maximum fan-in is 64 connections and the maximum fan-out is 4k connections. TrueNorth \cite{akopyan2015truenorth} has 4096 cores, and each core includes 256 Leaky Integrate-and-Fire (LIF) model neurons. Synapses, neurons, and axons are organized in the form of crossbars. 4096 cores are connected together through a 2D-mesh NoCs. Loihi \cite{davies2018loihi} is a digital neuromorphic chip developed by Intel. Each chip of Loihi has 128 neuromorphic cores and each core has 1024 neuromorphic units. Each core can simulate 130,000 LIF neurons and 1.3 billion synapses with a learning engine that supports on-chip training. 

\subsection{Mapping Tools of Neuromorphic Platforms}
Since the architecture of each neuromorphic platform is different, a dedicated toolchain is required to enable SNN to efficiently simulate on the neuromorphic platform. SpiNNaker \cite{furber2012overview} is a 2D toroidal mesh structure. PACMAN \cite{galluppi2012hierachical} was proposed to address SNN mapping on SpiNNaker. PACMAN uses a simulated annealing algorithm to search out the best partitioning scheme. But PACMAN only partitions the SNN model, which leads to spike congestion on the NoC. TrueNorth \cite{akopyan2015truenorth} also has their own mapping tool - {\itshape corlet} \cite{amir2013cognitive}. It uses the layout and routing optimization scheme in the traditional VLSI field for the mapping of logical SNNs to physical cores. SpiNemap \cite{balaji2019mapping} is proposed for the 2D-mesh architecture of Dynapse \cite{moradi2017scalable}. It divides the mapping process into two phases: partitioning and placement. They design a greedy Kernighan-Lin algorithm used in the partitioning phase and use the particle swarm optimization algorithm in the placement phase. For some neuromorphic platforms designed by new devices, \cite{lee2019system} \cite{xia2019memristive} were proposed to enable SNN to effective run on these neuromorphic platforms.

\section{toolchain}
\subsection{Overview}
\begin{figure}[ht]
  \centering
  \includegraphics[width=\linewidth]{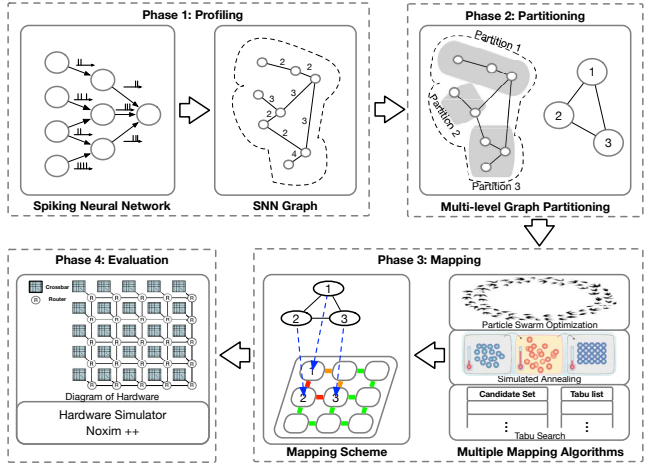}
  \caption{Overview of SNEAP.}
  \vspace{-0.5cm}
  \label{sneap}
\end{figure}

The toolchain we proposed maps SNN onto the NoC-based neuromorphic platform, is called SNEAP ( \underline{S}piking \underline{NE}ural network m\underline{AP}ping toolchain ). As shown in Figure \ref{sneap}, SNEAP consists of 4 phases: 1. {\itshape Profiling phase}: The topological structure of the trained SNN network and the behavior of neurons are extracted by the SNN network software simulator to form an undirected graph; 2. {\itshape Partitioning phase}: Partitioning divides the graph into multiple partitions based on the capability of the target neuromorphic platform. The multi-level partitioning algorithm is used to minimize spike communication among partitions; 3. {\itshape Mapping phase}: A selected algorithm is used to distribute these partitions to NoC of the target hardware, which minimizes the average-hop of all spikes on NoC of target hardware; 4. {\itshape Evaluation phase}: The mapping scheme is evaluated by NoC-based hardware simulator - Noxim++ \cite{balaji2019mapping}, so as to get key performance statistics.
\subsection{Profiling}
SNN software simulators (CARLsim \cite{chou2018carlsim}, Nest \cite{Gewaltig:NEST}, etc.) have been widely used by neuroscientists to precisely simulate the behavior of SNN. At present, most of SNN software simulators provide programming interfaces for developer to construct SNN. After construction, the developer can use attributes of the SNN to configure the SNN software simulator. These attributes include the number of neurons, neuron dynamic model, network topology and etc.

In this paper, we use CARLsim \cite{chou2018carlsim} to extract the connection information of the SNN and the behavior of spike. After we define the structure and connection scheme of SNN, we set the programming interface of CARLsim for simulation. When the simulation is finished, the log files of CARLsim are analyzed to generate graph with neurons as vertices and with synapses as edges between neurons. The weights of the edges are the number of spikes communicated on synapses. In addition, spike trace file can be obtained during the simulation. Each trace in spike trace file shows the specific behavior of each spike, and contains the ID of the source and destination neurons and firing time. Then we can perform partitioning and mapping on the SNN through the obtained graph and spike trace file. 

\subsection{Partitioning}
\label{sec:partition}
In our work, we propose to use a multi-level graph partitioning paradigm \cite{karypis1998multilevelk} to construct our partitioning tool. This tool solve SNN partitioning problem with the goal of minimizing the number of spikes between partitions. 

Partitioning problem can be transformed into $G(N,S)$ $\to$ $P(V,E)$.  
This is a classic graph partitioning problem. The graph partitioning problem is NP-complete problem.  Previously works use classic algorithms to solve the problem, such as particle swarm optimization (PSO) \cite{kennedy2010particle}, Kernighan-Lin (KL) \cite{kernighan1970efficient}, etc.  However, these approachs take a lot of time to find out the better partitioned SNN. We use a multi-level graph partitioning method to optimize the partitioning of large-scale SNNs. For the purpose of SNN partitioning, we introduce the following notations.
\vspace{-0.3cm}
\begin{table}[ht] 
  \begin{tabular}{lp{6.9cm} }
  $G(N,S)$ & = SNN graph with a set $N$ of vertices (neurons) and a set $S$ of edges (i, j) (synapses).\\
  $P(V,E)$ & = Partitioned SNN graph with a set $V$ of vertices (partitions) and a set $E$ of edges between partitions. \\
  $G_i(N_i, S_i)$& = The $i$-th level coarsening graph.\\
  $D_c[v]$      & = Partitioning vector in $c$-th level uncoarsening representing vertex $v$ belong to which partition.\\
  $B(v)$& = The union of the partitions that the vertices adjacent $v$ belong to.\\
  $ED[v]_b$& = External degree. For every $b$ $\in$ $B$($v$), $ED[v]_b$ is the sum of the weights of edges $(v,u)$ such that $D_c[u]$ = $b$.\\
  $ID[v]$& = Internal degree. $ID[v]$ is the sum of the weights of edges $(v,u)$ such that $D_c[u]$ = $D_c[v]$.
  \end{tabular}
\end{table}
\vspace{-0.3cm}

\begin{figure}[ht]
  \vspace{-0.3cm}
  \centering
  \includegraphics[width=\linewidth]{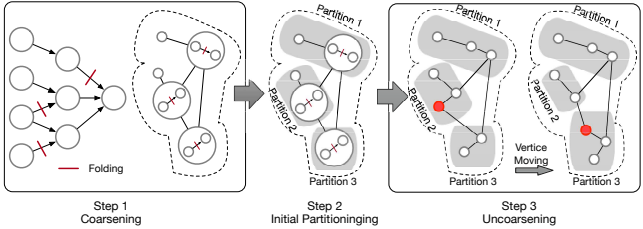}
  \vspace{-0.6cm}
  \caption{Multi-level graph partitioning diagram.}
  \vspace{-0.4cm}
  \label{mgp}
\end{figure}

Multi-level graph partitioning paradigm \cite{karypis1998multilevelk} consists of three steps (shown in Figure \ref{mgp}): {\itshape Coarsening, Initial partitioning, Uncoarsening}.  

{\itshape Coarsening} step is divided into multiple levels, and an original graph $G_0(N_0,S_0)$ is coarsened level by level. In the $i$-th level of coarsening, a set of vertices of $G_i$ is combined to form a single vertex of the next level coarser graph $G_{i+1}$. The vertices in graph $G_i$ are randomly selected. If a vertex $m$ is not folded yet, we fold a vertex $m$ with vertex $n$ such that the weight of the edge $(m, n)$ is maximum overall valid adjacent edges, which forms a vertex $v$ of graph $G_{i+1}$. We mark $m, n$ vertices as folded, and then repeat the above process until there is no more vertex that can be folded.

{\itshape Initial partitioning} step divides the graph $G_c$ generated by the coarsening step into $k$ partitions. The upper bound of the total vertex weight of each partition is decided by the number of neurons that can be accommodated in a neuromorphic core.  A vertex $m$ in graph $G_c$ is randomly selected to insert into partition $k$.  We search out an edge $(m, n)$ with the largest weight from the set of adjacent edges of partition $k$ and then insert vertex $n$ into partition $k$.  When a vertex inserts into partition $k$, the set of adjacent edges of partition $k$ is updated. We end inserting partition $k$ process if the total vertex weight of partition $k$ reaches the upper bound of partition $k$. Follow this process until the graph is divided into $k$ partitions.

{\itshape Uncoarsening} step, similar to the Coarsening step, is also divided into multiple levels. The partitioning $P_c$ of the coarser graph $G_c$ ($N_c$, $S_c$) is projected back to the original graph $G_0$. We use a global priority queue that stores the vertices according to their gains. Initially, all the vertices are scanned, and those whose sum of $ED$ is greater or equal to their $ID$ are inserted into the priority queue. In particular, let $v$ be such a vertex and $b$ $\in$ $B(v)$ such that $ED[v]_b$ is maximum in $B(v)$. We insert $v$ into the priority queue with a gain equal to $ED[v]_b-ID[v]$.  A vertex $v$ is selected from the global priority queue with the highest gains. We move vertex $v$ to partition $b$ that $ED[v]_b$ is maximum while satisfying the capacity of neuromorphic core. We continue moving vertices until $x$ vertex moves that have not decreased the sum of edge weights among partitions.  In that case, the last $x$ moves are undone. After such uncoarsening level by level, the optimized $k$ partitions are finally obtained.

Our proposed method use heuristics to quickly compress a large graph in the {\itshape Coarsening} step, so that the subsequent optimization steps will reduce the time consumption due to the large reduction in the size of the graph. Furthermore, since in the {\itshape Uncoarsening} step the single priority queue contains only vertices whose sum of $ED$ is greater or equal to their $ID$, this method has less powerful hill-climbing capabilities than the generalized KL \cite{kernighan1970efficient} that uses multiple priority queues and considers all the vertices.

\subsection{Mapping}

\begin{figure}[ht]
  \vspace{-0.4cm}
  \centering
  \includegraphics[width=\linewidth]{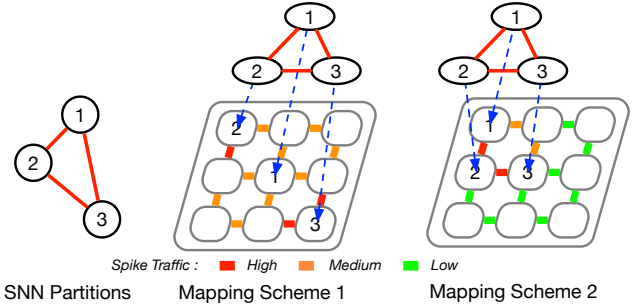}
  \vspace{-0.7cm}
  \caption{Congestion impact of different mapping schemes on neuromorphic platform.}
  \vspace{-0.4cm}
  \label{mapping}
\end{figure}

After the SNN is divided into multiple partitions, the placement of partitions on the neuromorphic platform also influences the latency and power consumption of the platform. As shown in Figure \ref{mapping}, different mapping schemes will change the communication behavior of spikes on the NoC, resulting in differences in power consumption and latency. In this paper, we implement three heuristic-based search algorithms to construct the mapping tool. The tool can find out the best mapping scheme that minimizes the spike latency and energy of the NoC-based neuromorphic platform. These search algorithms are Simulated Annealing algorithm (SA), Particle Swarm Optimization (PSO), and Tabu Search algorithm (Tabu) respectively.

The optimization objective of mapping could be latency and/or energy. However, evaluation of these metrics often requires using real hardware or hardware simulator, which leads to substantial time overhead and makes the entire search process unacceptable. As mentioned in section \ref{sub:average hop}, average hop is used to measure the latency and power consumption on the NoC. Compared with the above two metrics, average hop is easier to get \cite{lee2007chip}. Thus, instead of minimizing latency and energy consumption, we decide to minimize the average hop. Since we adopt XY routing algorithm in our neuromorphic platform, we propose a method of evaluating the average hop based on the XY routing algorithm, which reduces the time overhead caused by using real hardware or hardware simulator.
\vspace{-0.3cm}
\subsubsection{Mapping Algorithms}
We implement three heuristic-based algorithms (SA, PSO, Tabu) for finding a mapping scheme with the smallest average-hop. As shown in section \ref{sec:partition}, the partitioned SNN can be represented as a graph $P (V, E)$. The architecture of NoC-based neuromorphic platform can be considered as a graph $A(C,I)$, where $C$ is the set of neuromorphic cores and $I$ is the set of connections among these cores for a given interconnect topology. Mapping $M$ can be transformed into $M:$ $P(V,E)$ $\to$ $A(C,I)$. Mapping $M$ is represented by a matrix $m_{ij}$ $\in$ ${\{0,1\}}^{|C|\times|V|}$ , where $m_{ij}$ is defined as:
\begin{equation}
m_{ij}=
\left\{\begin{array}{ll}{1} & {\text { if partition } c_{i} \in C \text { is mapped to core } v_{j} \in V} \\
{0} & {\text { otherwise }}\end{array}\right.
\end{equation}
The optimization objective of our mapping phase is to find the mapping with the minimum average hop count $H$, i.e.
\begin{equation}
H_{min}  = min\{H(M_{i})|i\in 1,2,...,N\}
\end{equation}
Where $N$ is the number of evaluated mapping schemes.

The three algorithms use the same heuristic function (section \ref{sub:average hop}) to measure a candidate mapping scheme. The input and output format of the three algorithms are also the same. The input is a random initialized scheme. The output is the best scheme the algorithm can find within the given time limitation. They differ in choosing the next scheme from neighbors. Neighbors are possible schemes derived from the current scheme. For example, in the current scheme, all partitions have a corresponding core. Swapping any two partitions and their cores leads to a new scheme. These algorithms adopt different search strategies to find the best from the new schemes. SA allows the search forwarding to a less optimal orientation with a certain possibility, which is good for jumping out of local optimum. PSO is a population-based algorithm. Every particle in the population adapts according to both the best population history solution and the best personal history solution. Tabu uses a list, which is called tabu list, to record every history moves. Using this history information, Tabu can avoid dead loop and jump out of local optimum.
\vspace{-0.2cm}
\subsubsection{Algorithm for average hop evaluation.}
\label{sub:average hop}
In the current NoC-based neuromorphic platforms, the XY static routing algorithm is mainly adopted. The XY static routing algorithm can avoid deadlocks and is very simple to implement in hardware. Thanks to the static feature of the XY routing algorithm, hop distance that spike traverses can be calculated directly without using hardware simulation. Based on this, we proposed a algorithm that can directly calculate the average hop.

We formalize the algorithm for average hop evaluation as Algorithm \ref{alg:evaluation}. First, we extract the communications between partitions from the spike trace. Then, we traverse the communication between any two partitions and calculate the distance between cores whose partitions are mapped. Finally, we multiply the distance by the corresponding total amount of communications to get average hop.
\vspace{-0.6cm}
\begin{algorithm}[H]
  \caption{Average Hop Evaluation Algorithm}\label{alg:A}
  \begin{algorithmic}[1]
  \State Input: the partitions $ (p_{1},p_{2},...,p_{n}) $, the number of cores $m$, the mapping option $ M $,the source core $s$,the destination core $d$, spike trace.
  \State Output: average hop H.
  \State $trace\ length\gets spike\ trace$
  \State $C_{n*n} \gets zero\,matrix$ // communications between partitions
  \For {spike in spike trace}
  \State $time\,step,neuron_{source},neuron_{destination} \gets $spike
  \State $neuron_{source}\in p_i,neuron_{destination}\in p_j$
  \State add a communication to $C(p_i, p_j)$
  \EndFor
  \For {a in partitions}
  \For {b in partitions}
  \State $s \gets M(a)$, $d \gets M(b)$
  \State $(x,y)_{core} \gets $get coordinate()
  \State $hop\,distance \gets |s_{x}-d_{x}|+|s_{y}-d_{y}|$
  \State $H \gets \sum_{a=0}^{n}\sum_{b=0}^{n} hop\ distance*C(a,b)\div trace\ length$
  \EndFor
  \EndFor
  \State \textbf{return} $ H $
  \end{algorithmic}
  \label{alg:evaluation}  
\end{algorithm}

\section{Experiment Setup}
\subsection{Experiment platform}
The experimental platform was constructed following two simulators and two tools. 

Two simulators are SNN software simulator - CARLsim \cite{chou2018carlsim} and hardware simulator - Noxim++ \cite{balaji2019mapping}. CARLsim is a GPU-accelerated software SNN simulator that can be used to train and test SNN networks. The behavior of spike can be analyzed from the log file of CARLsim. Noxim++ is a trace-driven and cycle-accurate NoC simulator. Noxim++ is an extension version based on Noxim \cite{catania2018improving}. Noxim++ is used to simulate the execution of SNN on real NoC-based hardware, so as to evaluate key performance statistics of NoC, such as average hop, delay, and power consumption. 

Two tools are partitioning tool and mapping tool. For the partitioning tool, we reference the Metis \cite{karypis1998multilevelk} with a python interface to implement it, including all key components of the multi-level partitioning paradigm. The mapping tool mainly contains three heuristic algorithms (SA, PSO, Tabu) and a component of evaluating average hop. Combined with the average hop evaluation component, these algorithms are used to search for the best mapping on the NoC-based neuromorphic platform.

Our experiment uses the hardware configuration of 5x5 2D-mesh NoC, and neuromorphic core adopts crossbar structure. Every crossbar can accommodate at most 256 neurons, meaning that a crossbar sends at most 256 spikes per time step.

All experiments were performed on i7-7700, 16GB RAM, and NVIDIA GTX1060 GPU, Ubuntu 16.04.
\subsection{Evaluated SNNs}
Table \ref{tab} provides a set of SNNs used to evaluate our proposed toolchain. These five SNNs have different topologies, including variety of depth and width of SNN layers and different connectivity-scheme.
\begin{table}[H]
  \vspace{-0.3cm}
  \caption{Evaluated SNNs. The number in the first column represents the number of neurons of the SNN.}
  \vspace{-0.3cm}
  \begin{tabular}{|c|c|c|}
    \hline SNN Name&Network Topology&Spikes\\
    \hline Smooth\_320\cite{chou2018carlsim}&Feedfoward, 2 layer&175124\\
    \hline Smooth\_1280\cite{chou2018carlsim}&Feedfoward, 2 layer&981808\\
    \hline MLP\_2048\cite{diehl2015unsupervised}&Feedfoward, 2 layer&15905792\\
    \hline Edge\_5120\cite{chou2018carlsim}&Feedfoward, 3 layer&4570546\\
    \hline Random\_6212\cite{chou2018carlsim}&Feedfoward, 3 layer&51756245\\
    \hline
  \end{tabular}
  \vspace{-0.3cm}
  \label{tab}
\end{table}

\subsection{Metrics for evaluation}
We evaluate all three mapping methods in terms of the following metrics for every SNNs.

{\itshape Energy consumption on the NoC} : This is the overall energy consumed by spikes communication on the NoC.

{\itshape Average latency} : This is the delay experienced by spikes before reaching their destination and averaged overall spikes.

{\itshape Congestion Count} : Beside latency and energy consumption, one essential metric that we get from the toolchain is congestion count, which reflects the degree of congestion on the NoC.
\begin{equation}
  Congestion_{Count} = \sum_{t=0}^{n}C_{t}
\end{equation}
During each time step $ t $, congestion is defined as the number of spikes exceed the mesh edge's load. The spikes that exceed the load cannot be transmitted at this time step, whose number is $ C_{t} $.

{\itshape Edge Variance :}
Same as Congestion Count, edge variance is used to reflect the degree of congestion and the load distribution on the NoC. We can get the total hop numbers of every edge on mesh network with XY static routing algorithm.
Supposed there are $n$ edges on mesh, $ e_{i} $ represent edge-$ i $'s total hop numbers after all time steps.
\begin{equation}
  Edge = (e_{1},e_{2},...,e_{n}) 
\end{equation}
\begin{equation}
  Edge_{Var} = Var(Edge) 
\end{equation}

\section{Results and Discussion}
In this section, we compare SNEAP with some state-of-the-art methods proposed by SpiNeMap and SCO \cite{lee2019system}. SpiNeMap uses SpiNeCluster to partition SNNs into clusters to minimize the total number of spikes among the clusters and SpiNePlacer to optimize the placement of clusters to crossbars of the neuromorphic hardware to minimize energy consumption and latency. SCO uses its framework to balance the utilization of crossbars in the hardware. We summarize the improvements of our method against SpiNeMap and SCO. We now describe these results in detail.
\vspace{-0.1cm}
\subsection{Partitioning Performance}
\begin{figure}[ht]
  \centering
  \includegraphics[width=8.7cm]{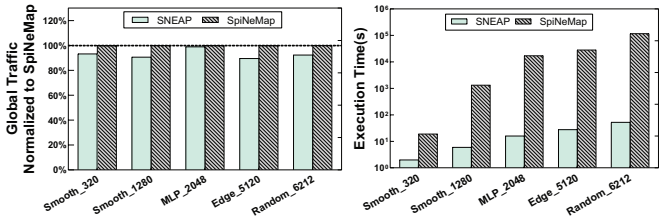}
  \vspace{-0.8cm}
  \caption{Performance in partitioning phase.}
  \label{partitioning}
\end{figure}

In Figure \ref{partitioning}, we compare the global traffic (the number of spikes among partitions) and the execution time of each SNNs under different methods normalized to SpiNeMap. Compared with SpiNeMap, SNEAP has a 890$\times$ reduction in execution time. The cause of this reduction is that the heuristic algorithm is used to compress a large graph quickly during the partitioning phase, so that the subsequent optimization process will reduce the time consumption. SNEAP has 8\% fewer average the number of spikes among partitions than SpiNeMap. These improvements  are as a result of the optimization algorithm of SNEAP, which is good for jumping out of local optimum.
\vspace{-0.3cm}
\vspace{-0.5cm}
\subsection{Mapping Algorithms Comparison}
\begin{figure}[ht] 
  \vspace{-0.3cm}
  \centering
  \includegraphics[width=7cm]{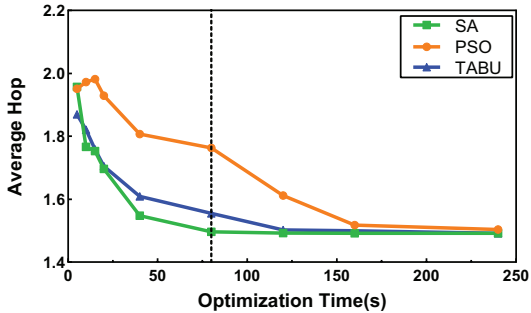}
  \vspace{-0.3cm}
  \caption{Comparison of convergence speed.}
  \vspace{-0.3cm}
  \label{alg} 
\end{figure}

As shown in Figure \ref{alg}, we evaluate the convergence time of three algorithms (SA, PSO, Tabu) and then get the relationship between average hop and time consumed. We also performed the same analysis on the other four types of SNN, and the results are similar. Because SA can search the best results in the shortest time in this type of optimization problem, so in this paper we use SA to find out the best mapping scheme.

\begin{figure}[ht]
  \vspace{-0.3cm}
  \centering
  \includegraphics[width=8.7cm]{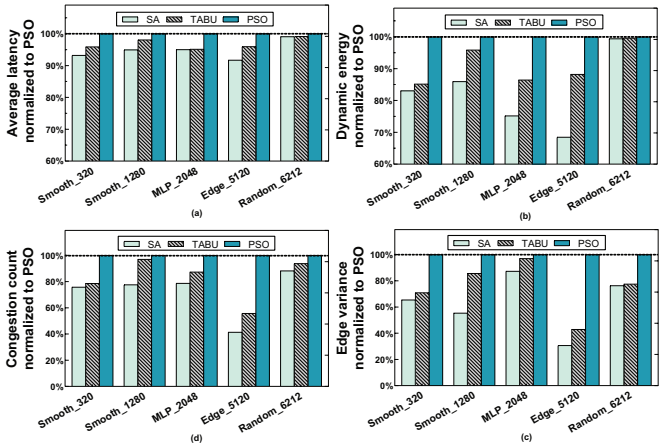}
  \vspace{-0.7cm}
  \caption{Evaluate various algorithm in mapping phase.}
  \vspace{-0.3cm}
  \label{al:mapping}
\end{figure}

Figure \ref{al:mapping} shows average latency, dynamic energy, congestion count, and edge variance on the mapping phase under different heuristic algorithms normalized to PSO proposed by SpiNeMap. As can be seen from Figure 6, SA results in about 1\% to 8\% and average 3\% reduction in average latency, almost 2\% to 33\% and average 16\% reduction in dynamic energy, nearly 15\% to 63\% and average 28\% in edge variance and approximately 12\% - 61\% and average 25\% in congestion count compared with other algorithms. In conclusion, SA can find the best mapping with lower energy and latency than other algorithms within a certain time period.

\subsection{Overall Toolchain Results}

\subsubsection{Average latency}
Figure \ref{total}(a) gives shows the average latency of overall spikes on the NoC under different method normalized to SpiNeMap. The statistic shows that compared with SpiNeMap and SCO, SNEAP has a great reduction in all of SNN cases. SNEAP results in average 51\% lower than the SpiNeMap and 88\% lower than SCO. These improvements are because of the optimization objective of SNEAP. SNEAP adopts objective to minimize the total number of spikes among the partitions and average hop. In addition to optimization objective, optimization algorithms are also better, which good for jumping out of local optimum. 

For the case of the largest SNN Random\_6212, SNEAP achieves 92\% lower average latency than SpiNeMap. While for the other case such as MLP\_6212, SNEAP only achieves 8\% lower average. The cause of this consequence is different connectivity-scheme between MLP\_2048 and Random\_6212. Compared to random connect (Random\_6212), full connect (MLP\_2048) has less optimizable space in the whole toolchain.
\vspace{-0.2cm}
\subsubsection{Energy}

Figure \ref{total}(b) gives the dynamic energy of the NoCs under different method normalized to SpiNeMap. Since all experiments are based on 5x5 2D mesh structure, static energy is always a constant. Consequently, we use the dynamic energy to evaluate the energy consumption of NoCs. Compared with other methods, SNEAP has the lowest energy consumption. SNEAP results in average 23\% lower than the SpiNeMap and 31\% lower than SCO.

The improvement is due to the multi-level partitioning algorithm, which outperforms the greedy KL algorithm proposed by SpiNeMap. Fewer spikes communicated among the partitions, lower dynamic energy consumption.

\vspace{-0.2cm}
\subsubsection{Congestion}

\begin{figure}[ht]
  \vspace{-0.2cm}
  \centering
  \includegraphics[width=8.7cm]{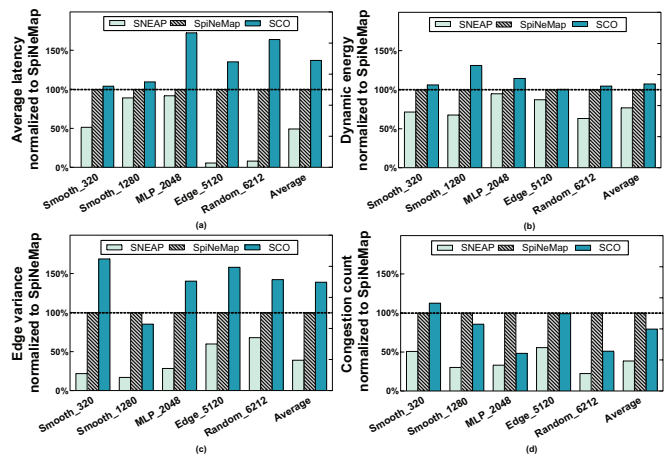}
  \vspace{-0.7cm}
  \caption{Overall Results.}
  \vspace{-0.4cm}
  \label{total}
\end{figure}

In Figure \ref{total}(c), we report the edge variance of the NoCs under different methods normalized to SpiNeMap. As shown in Figure \ref{total}(c), SNEAP has the lowest edge variance  of all our evaluated methods. For SpiNeMap, SNEAP has an average 61\% reduction. For SCO, an average reduction is 1$\times$. This reduction is due to the partitioning algorithm of SNEAP, which may adopt a non-optimal solution to jump out of local optimal compared with greedy KL used by SpiNeMap. This indirectly leads to a balanced distribution of spikes on the NoCs. 

Figure \ref{total}(d) presents the congestion count of the NoCs under different methods normalized to SpiNeMap. The results of the congestion count are similar to that of the edge variance. The more balanced mapping of spikes can effectively reduce the congestion count on the NoC.
\vspace{-0.2cm}
\subsubsection{Execution time of toolchains}
\begin{figure}[ht]
  \vspace{-0.3cm}
  \centering
  \includegraphics[width=6.5cm]{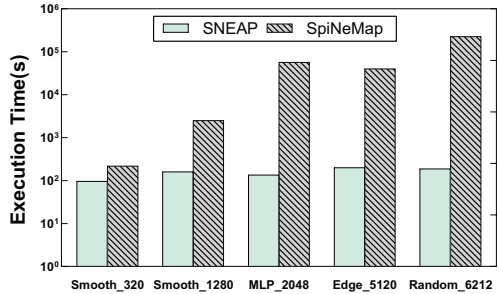}
  \vspace{-0.3cm}
  \caption{Execution time of toolchains.}
  \vspace{-0.5cm}
  \label{time}
\end{figure}

In Figure \ref{time}, we illustrate the end-to-end execution time under different toolchains. SNEAP achieves 418$\times$ lower average execution time than SpiNeMap. The causes behind this are that during the partitioning phase SNEAP has a reduced amount of execution time compared to SpiNeMap and that in mapping phase SA converges faster than PSO.

\section{conclusion \& future work}
This paper presents a fast and efficient toolchain - SNEAP to map the large-scale SNN onto the NoC-based neuromorphic platform. SNEAP completes the entire mapping process in four phases: Profiling, Partitioning, Mapping, Evaluation. In the profiling phase, we use the SNN software simulator to extract the essential information of SNN such as topology and the behavior of spike. By using this information, we construct the undirected graph of SNN and generate spike trace files. In the partitioning phase, we use a multi-level graph partitioning method to quickly divided the graph of SNN into multiple SNN partitions. Our objective is to minimize the number of spikes between partitions. In the mapping phase, we use the heuristic-based algorithm (SA) to map optimized SNN partitions on the physical processing unit in hardware. Combining the optimization in the partitioning phase, heuristic-based mapping algorithm optimizes the energy consumption and spike latency on the NoC-based neuromorphic platform. Using five SNNs, we show that our toolchain can achieve 418$\times$ reduction in end-to-end execution time, and reduce average energy consumption by 23\% and average spike latency by 51\%, compared to SpiNeMap. In the future, the toolchain is to support mapping optimization during the learning process of SNNs. In the learning process of SNNs, the topology of the SNNs changes dynamically, which brings challenges to the partitioning and mapping tasks of the toolchain.

\bibliographystyle{unsrt}
\bibliography{GLSVLSI_arXiv} 


\end{document}